\documentclass[fleqn,usenatbib]{mnras}

\usepackage{newtxtext,newtxmath}

\usepackage[T1]{fontenc}

\DeclareRobustCommand{\VAN}[3]{#2}
\let\VANthebibliography\thebibliography
\def\thebibliography{\DeclareRobustCommand{\VAN}[3]{##3}\VANthebibliography}

\usepackage{graphicx}	
\usepackage{amsmath}	
\usepackage{xcolor}
\usepackage{pdfcolfoot}

\newcommand{\bol}[1]{\boldsymbol{#1}}

\title[Subhalo profiles with machine learning]{A deep learning model for the density profiles of subhaloes in IllustrisTNG}

\author[Lucie-Smith, Despali \& Springel]{%
Luisa Lucie-Smith$^{1}$\thanks{E-mail: luisals@mpa-garching.mpg.de},
Giulia Despali$^{2,3,4}$
and Volker Springel$^{1}$
\vspace*{1mm}\\%
$^{1}$Max-Planck-Institut f{\"u}r Astrophysik, Karl-Schwarzschild-Str. 1, 85748 Garching, Germany\\%
$^{2}$ Dipartimento di Fisica e Astronomia "Augusto Righi", Alma Mater Studiorum Università di Bologna, via Gobetti 93/2, I-40129 Bologna, Italy \\%
$^{3}$ INAF-Osservatorio di Astrofisica e Scienza dello Spazio di Bologna, Via Piero Gobetti 93/3, I-40129 Bologna, Italy\\%
$^{4}$  INFN-Sezione di Bologna, Viale Berti Pichat 6/2, I-40127 Bologna, Italy\\%
}

\date{Accepted XXX. Received YYY; in original form ZZZ}

\pubyear{2015}

\begin{document}
\label{firstpage}
\pagerange{\pageref{firstpage}--\pageref{lastpage}}
\maketitle

\begin{abstract}
We present a machine-learning-based model for the total density profiles of subhaloes with masses $M \gtrsim 7\times 10^8\,h^{-1}{\rm  M}_\odot$ in the IllustrisTNG100 simulation. The model is based on an interpretable variational encoder (IVE) which returns the independent factors of variation in the density profiles within a low-dimensional representation, as well as the predictions for the density profiles themselves. The IVE returns accurate and unbiased predictions on all radial ranges, including the outer region profile where the subhaloes experience tidal stripping; here its fit accuracy exceeds that of the commonly used Einasto profile. The IVE discovers three independent degrees of freedom in the profiles, which can be interpreted in terms of the formation history of the subhaloes. In addition to the two parameters controlling the normalization and inner shape of the profile, the IVE discovers a third parameter that accounts for the impact of tidal stripping onto the subhalo outer profile; this parameter is sensitive to the mass loss experienced by the subhalo after its infall onto its parent halo. Baryonic physics in the IllustrisTNG galaxy formation model does not impact the number of degrees of freedom identified in the profile compared to the pure dark matter expectations, nor their physical interpretation. Our newly proposed profile fit can be used in strong lensing analyses or other observational studies which aim to constrain cosmology from small-scale structures.
\end{abstract}




\section{Introduction}
A key prediction of the cold-dark-matter (CDM) paradigm is that dark matter haloes are not smooth: they are filled with a large number of self-bound substructures \citep{Diemand2008, Springel2008}, known as subhaloes, wherein galaxy formation takes place \citep[see e.g.][]{ZavalaFrenk2019}. The abundance and density structure of subhaloes are sensitive probes of the fundamental nature of dark matter; for example, alternative dark matter models including warm dark matter and ultra-light scalar field dark matter suppress subhalo abundances on small scales \citep[e.g.][]{Maccio2010, Schive2014}. 

The existence of subhaloes is a natural consequence of hierarchical structure formation: haloes are built up both through mergers with smaller structures that survive as subhaloes within the larger host and by some degree of smooth mass accretion \citep{Wang2011}. After their initial radial fall into the host halo, these subhaloes are subject to a gradual loss of energy due to dynamical friction and tidal forces in the host halo \citep{BinneyTremaine2008, MovandenBosch2010}. Subhaloes survive until dynamical friction causes them to sink into the host’s centre, or they are completely disrupted by tidal forces which unbind their particles and mix them into the larger host halo. The population of surviving subhaloes will exhibit changes to their density profiles compared to isolated haloes, as matter in the outer region of the subhaloes is stripped away due to tidal forces.

The most direct method to search for the ubiquitous small-mass dark haloes beyond our galaxy is strong gravitational lensing. The distortions of images observed in lensing are due to gravity only, thus allowing us to measure the total mass of the deflectors -- both luminous and dark -- directly. Subhaloes can be detected by the effect they have on the flux ratios of multiply-imaged quasars \citep{flux_ratios2, flux_ratios1} or by the perturbations they cause to the surface brightness of extended arcs in strong galaxy-galaxy lens systems \citep{vegetti2010a, vegetti2012, hezaveh2016, Despali22,ORiordan2023,Nightingale2024}. Flux ratio anomalies are sensitive to the integrated effect of the population of perturbers, i.e. subhaloes in the main lens or dark isolated haloes along the line-of-sight. In extended arcs, a detailed lens modeling can be used to infer the mass of individual perturbers, which range from $1.9 \times 10^8$ to $2.7 \times 10^{10}\,{\rm M}_\odot$ for current detections. Lensing can robustly infer the mass within the Einstein radius (in practice the location where the lensed images appear), while the total mass depends on the underlying assumptions of (sub)halo distribution and density profiles, typically derived from numerical simulations. Recent works \citep{Minor21, Sengul22} found that the concentrations of previously detected subhaloes are exceptionally high when their profile is modelled with Navarro-Frenk-White (NFW; \citealt{NFW}) profiles, making them 2$\sigma$ outliers of the CDM model. Alternative dark matter models, such as self-interacting dark matter, have been proposed as a solution to this tension \citep{Nadler2023}. Another possibility is that the detected perturbers are subhaloes which experienced tidal stripping, whose profiles therefore deviate significantly from the standard NFW model \citep{Heinze24}.
In addition to a model of the number density of subhaloes \citep{Despali2017}, a detailed understanding of the density profiles of dark matter substructures is thus key to inferring their correct mass from observational data. 

In order to derive constraints on the (sub)halo mass function by comparing observations of gravitational lensing with theoretical predictions, it is essential to understand the density distribution of substructures. For example, while isolated dark-matter haloes follow NFW density profiles to a good approximation, this need not be the case for subhaloes. These are identified in numerical simulations as secondary density peaks within the main halo, their density profiles differ from the NFW profile, particularly in the outskirts, and their virial mass (one of the two NFW parameters) is typically ill-defined. 

Our goal is to provide a new model for the density profiles of subhaloes, which includes the effect of stripping and baryonic physics. To fully describe the variety of subhalo profiles, one needs a functional form with more free parameters than the classic NFW profile -- which is well described by the density normalisation and scale radius. Previous works have already derived analytic parametrizations of subhalo profiles, such as the Einasto profile \citep{Einasto1965}, the truncated NFW profile \citep{Baltz2009, Green2019}, or more complex profiles \citep{Stoehr2006, DiCintio2013, Heinze24}, thus needing three or more (often degenerate) parameters. However, choosing a specific functional form can limit the performance of the model and it is not trivial to determine how many parameters would provide the best fit, given that the parameters are usually highly correlated. \citet{Heinze24} have recently shown that individual subhalo density profiles depend on multiple physical properties of the subhalo (distance from the host halo, concentration, infall time or baryonic fraction) at the same time, and that these properties are correlated. In this paper, we use machine learning to leverage these issues and determine the number of degrees of freedom required to describe the density profiles of realistic subhaloes from hydrodynamical simulations. 

\citet{LucieSmith2022} tackled a similar problem in the context of field dark matter haloes in gravity-only simulations. They modelled the spherically-averaged density profiles of field dark matter haloes using an \textit{interpretable variational encoder} (IVE); the model not only provides accurate predictions taking advantage of the flexibility of machine-learning models, but also generates a compact, low-dimensional latent representation that is equivalent to the independent degrees of freedom in the output of interest. They showed that three degrees of freedom are required (and sufficient) to model field halo density profiles; two of these resemble the NFW mass and concentration parameters, while the additional latent contains information about the dynamical, unrelaxed component of the haloes similar to the splashback effect. In \citet{LucieSmith2023}, they exploited the latent representation further by going beyond its original training task. They demonstrated that the latent representation discovered by the neural network from $z=0$ data carries memory of the evolution history of the haloes, thus shedding light onto the origin of the density profiles degrees of freedom in terms of the haloes' accretion histories.

In this work, we extend their work to construct a new data-driven model for density profiles of substructures in both gravity-only and hydrodynamical simulations. The paper is structured as follows. We describe the set of simulations used in this work in Sec.~\ref{sec:sims}, and the construction of the training data in Sec.~\ref{sec:data}. In Sec.~\ref{sec:IVE}, we present an overview of our framework, including details on the machine-learning model and the training procedure. We show the predictive performance of our model in Sec.~\ref{sec:pred} and then move to interpreting the latent representation discovered by the neural network in Sec.~\ref{sec:latents}. We then show the relation between the latent and physical parameters in Sec.~\ref{sec:physicalparams}. We draw our final conclusions in Sec.~\ref{sec:conclusions}.

\section{Simulations}
\label{sec:sims}
\begin{figure*}
    \centering
	\includegraphics[width=0.9\textwidth]{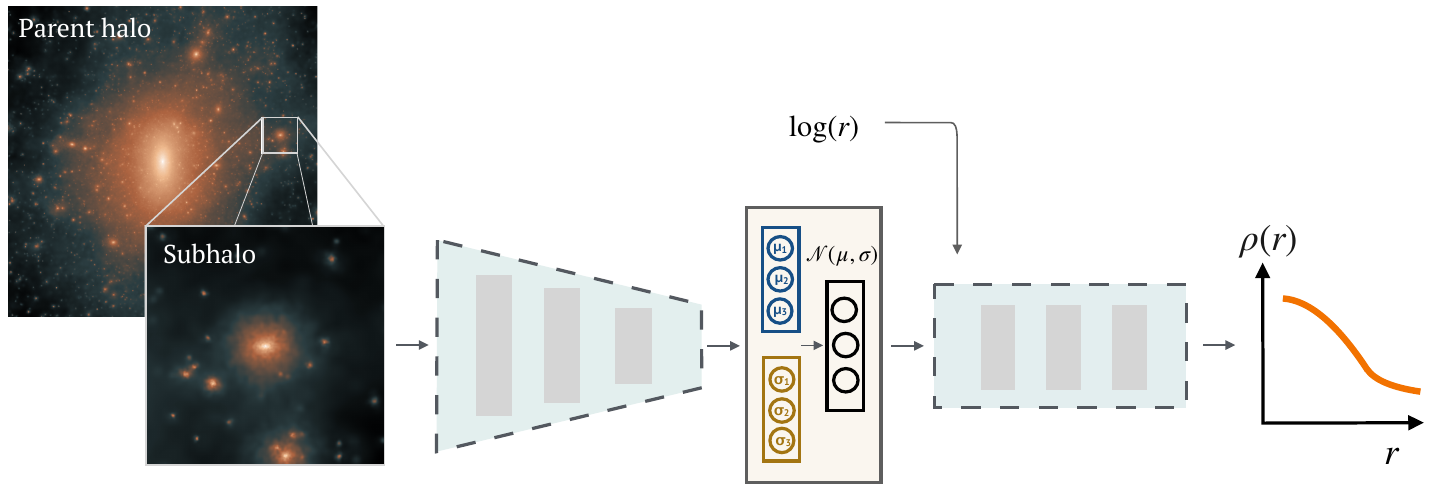}
    \caption{The IVE takes as input a cubic sub-region of the simulation centred around the subhalo, and outputs the spherically-averaged density $\rho(r)$ as a function of (log of) $r$ (i.e. the subhalo density profile). The input is first compressed by the encoder into a low-dimensional latent space, where each latent is a Gaussian distribution. The decoder then maps samples from the latent space and a given value of $r$ to the spherically-averaged total density $\rho(r)$; the latter flattens out towards the virial radius where the background dominates. In this illustration, the latent space is three-dimensional. Note that the inputs to the encoder are in 3D but shown as 2D projections in this illustration.}
    \label{fig:architecture}
\end{figure*}

The training data for the machine-learning model were constructed using the IllustrisTNG simulations \citep{Springel2017, Pillepich2017}. The IllustrisTNG project is a suite of state-of-the-art cosmological magnetohydrodynamical simulations, which include a comprehensive model for galaxy formation physics and adopt a Planck cosmology \citep{Planck2016}. In particular, we make use of the IllustrisTNG-100-1 (TNG100 hereafter) simulations which provides the best compromise between volume and resolution for probing the halo (and subhalo) mass range we wish to consider. The simulation has a volume of $(75 \,h^{-1} \rm{Mpc})^3$ traced by $2 \times 1820^3$ resolution elements, with a dark matter mass of $m_\mathrm{DM} = 5.1 \times 10^6 \,h^{-1} \mathrm{M_{\odot}}$ and a baryon mass of $m_\mathrm{baryon} = 9.4 \times 10^5 \,h^{-1} \mathrm{M_{\odot}}$. We also employ its dark-matter-only counterpart for comparison, TNG100-1-Dark (TNG100-Dark hereafter), which was evolved with the same initial conditions and the same number of dark matter particles with $m_\mathrm{DM} = 6 \times 10^6 \,h^{-1} \mathrm{M_{\odot}}$. The dark matter softening length is $\epsilon = 0.74 \, \mathrm{kpc}$. 

The dark matter haloes in TNG100 (and TNG100-Dark) were identified at $z=0$ by applying the substructure finder {\small SUBFIND} \citep{Springel2001} to friends-of-friends haloes found with a linking length of 0.2. This determines subhaloes as locally overdense groups of particles and cells that are gravitationally bound. We consider subhaloes that live within host haloes of mass $M_\mathrm{host} \geq 10^{11} \, h^{-1} \mathrm{M}_{\odot}$. Out of those, we further select subhaloes with a total number of resolution elements $N_\mathrm{p} \ge 150$ and a virial radius $R_\mathrm{vir} \leq 200 \,h^{-1} \mathrm{kpc}$. The virial radius $R_\mathrm{vir}$ is here defined as the smallest radius enclosing the total bound mass of the subhalo. We make the same cuts when selecting the subhaloes in TNG100 and TNG100-Dark.

In addition to the halo and subhalo catalogues, we also make use of  merger trees constructed with the {\small SUBLINK} \citep{Rodriguez2015} algorithm, which are part of the data released with the TNG simulations \citep{Nelson2019}. These merger trees are constructed at the subhalo level, by identifying progenitors and descendants of each subhalo.  Firstly, descendant candidates are identified for each subhalo as those subhaloes in the following snapshot that have common particles with the subhalo in question. These are then ranked based on the binding energy of the shared particles, and the descendant of a subhalo is defined as the candidate with the highest score. Knowledge of all the subhalo descendants, along with the definition of the first progenitor, uniquely determines the merger trees. It is thus possible to walk the tree backward in time to determine when each subhalo formed as an independent structure, and when it then fell into the host halo.

\section{Training data}
\label{sec:data}

\subsection{Outputs: Spherically-averaged density profiles}

We used the $z=0$ snapshot of the simulation to assign to each subhalo its ground-truth spherically-averaged density profile. We use the density profile of the cold dark matter for the TNG100-Dark subhaloes, and the total density profile including contributions from cold dark matter, gas, stars and black holes when using TNG100 ones. We choose to model the total density profile since observational probes, such as strong lensing and other direct profile measurements, are sensitive to the total mass within (sub)structures. We adopt as centre the position of the particle with the minimum gravitational potential energy. For every subhalo, we computed its density profile by evaluating the density within 24 bins in radius, logarithmically-spaced in the range $r \in [1\,h^{-1} \mathrm{kpc}, R_\mathrm{vir}] $, where $R_\mathrm{vir}$ is the virial radius of the subhaloes assuming $M_\mathrm{vir}$ to be the total bound mass. 

The density $\rho(r)$ is computed using all particles at distance $r$ from the subhalo centre, not just those gravitationally bound to the halo as identified by the halo finder. This choice is motivated by the fact that lensing observations are sensitive to the total mass distribution, and not just that coming from particles gravitationally bound to the subhalo according to the halo finder. Subhalo profiles are tidally truncated by an amount that depends on their intrinsic properties and history inside the main halo. This has motivated the use of Einasto \citep{Einasto1965}, ``pseudo-Jaffe'' \citep{Munoz2001} or truncated NFW \citep{Baltz2009} profiles to describe the density distribution of subhaloes; however, these models only provide reasonable fits when considering only particles that are gravitationally bound to the halo. Here, we aim to describe the entire density field around subhalo centres, similarly to what was done for haloes in \citet{LucieSmith2022}. In practice, this means that our density profiles are not visually truncated but rather flattened at large radii where the background density of the halo starts dominating, as we will discuss further below. 

\subsection{Inputs: 3D density cubes}
The inputs were generated from the 3D density field, $\rho(\mathbf{x})$, at $z=0$. More specifically, the input for each subhalo is given by $\log [\rho(\mathbf{x})/\bar{\rho}_{\rm m} + 1]$ in a cubic sub-region of the full simulation of size $L_\mathrm{sub-box} = 200 \,h^{-1} \mathrm{kpc}$ and resolution $N_\mathrm{sub-box} = 101^3$, centred on the subhalo centre \footnote{An alternative choice of input would be the 1D spherically averaged density profile. However, we opt for the more generalizable choice of a network architecture utilizing the 3D density field as input, which allows for straightforward extensions to other halo observables, such as halo triaxiality or accretion histories, which require the entire 3D density field as input.}. $\bar{\rho}_{\rm m}$ is the mean matter density of the universe. When training on the TNG100-Dark subhaloes, the density field is that from the cold dark matter component only; when training on TNG100, we instead use the total density field which includes contributions from cold dark matter, gas, stars and black holes (when present). All subhaloes, independently of their size and mass, have input sub-boxes of the same size and resolution. The choice of volume and resolution of the input sub-boxes was made to ensure the IVE has access to the relevant scales: the voxel size, $l \sim 2\,h^{-1} \mathrm{kpc}$, matches the smallest radial value of the profile, and the sub-box size is 2 times larger than the virial radius of more than $99\%$ of the subhaloes.  Depending on the location of the subhalo with respect to the parent halo, and on the overall size of the halo, the input sub-box will cover different fractions of the volume of the parent halo. The network is not explicitly provided with information about the parent halo of each subhalo; however, the network implicitly has access to information about the parent halo's properties which affects the subhalo environment via the input density field. The density field was constructed from the positions of particles in the simulation using a smoothed-particle hydrodynamics (SPH) procedure, and then evaluated at each voxel of the cubic sub-box.

\section{An interpretable variational encoder}
\label{sec:IVE}
We used an IVE to model the density profiles of subhaloes giving as input the ``raw'' 3D density field surrounding each subhalo centre.  The IVE architecture used in this work was first developed in \cite{LucieSmith2022}, and has two main components: the encoder, mapping the 3D density field to a low-dimensional latent representation, and the decoder, mapping the latent representation and an additional input -- the query radius $\log(r)$ --  to the output profile $\log[\rho(r)]$. By design, the latent space contains all the information required by the model to predict the output $\rho(r)$ as a function of any query value $\log(r)$; in other words, it captures all the information in the inputs about the density profile. An illustration of the model is shown in Fig.~\ref{fig:architecture}.

The encoder is a 3D convolutional neural network with parameters $\phi$ that maps the input density $\delta(\bol{x})$ to a multivariate distribution in the latent space $p_{\phi}(\bol{z} | \delta(\bol{x}))$. We choose the latent representation to be a set of independent Gaussians, $p_{\phi} (\bol{z} | \bol{x}) = { \prod_{i=1}^L}  \mathcal{N}(\mu_i(\bol{x}), \sigma_i(\bol{x}))$, where $L$ is the dimensionality of the latent space; under this assumption, the encoder maps the inputs $\delta(\bol{x})$ to the vectors of means $\mu = {\mu_i, .., \mu_L}$ and standard deviations $\sigma = {\sigma_i, .., \sigma_L}$. The dimensionality of the latent space $L$ is a hyperparameter of the network that must be set prior to training. The typical strategy is to train the IVE using increasing values of $L$ until the accuracy no longer increases; the smallest $L$ providing the highest possible accuracy is equivalent to the underlying dimensionality of the output of interest. The decoder of the IVE consists of another neural network model with parameters $\theta$ that maps a sampled latent vector $z \sim p_{\phi} (\bol{z} | \delta(\bol{x}))$ and a value of the query $\log(r)$ to a single predicted estimate for $\log[\rho_\mathrm{pred}(r)]$.

\subsection{The loss function}
Training the IVE involves optimizing the parameters of the encoder $\phi$ and of the decoder $\theta$ such that the loss function is minimized. The loss function is given by \citep{Higgins2017},
\begin{equation} 
\mathcal{L} = \mathcal{L}_\mathrm{pred}( \bol{\rho}_\mathrm{true}, \bol{\rho}_\mathrm{pred} )  + \beta \, \mathcal{D}_\mathrm{KL}[p_{\phi}(\bol{z} | \bol{x}); q(\bol{z})],
\label{eq:loss}
\end{equation}
where the first term measures the predictive accuracy of the model and the second is the Kullback-Leibler (KL) divergence \citep{Kullback1951} between the latent distribution returned by the encoder $p_{\phi}(\bol{z} | \bol{x})$  and a prior distribution over the latent variables $q(\bol{z})$. The parameter $\beta$ weighs the KL divergence term with respect to the predictive term, and must be carefully optimized. We took the predictive term to be the mean squared error loss, 
\begin{equation}
\mathcal{L}_\mathrm{pred} = \frac{1}{N}\sum_{i=1}^{N} \left( \log_{10} \rho_{i,\mathrm{true}} - \log_{10} \rho_{i,\mathrm{pred}} \right)^2,
\label{eq:losspred}
\end{equation}
where $N$ is the training set size. Assuming a set of independent unit Gaussian distributions as the prior over the latent variables $q(\bol{z)}$, the KL divergence term takes the closed form,
\small
\begin{equation}
\mathcal{D}_\mathrm{KL}(\mathcal{N}(\mu_{\bol{z}}, \sigma_{\bol{z}}); \mathcal{N}(0, 1)) = -\frac{1}{2}\sum_{i=1}^L \left[ 1 + 2\log \sigma_i - \mu_i^2 - \sigma_i^2 \right] ,
\label{eq:KL}
\end{equation}
\normalsize
where $L$ is the dimensionality of the latent space. The role of the KL term in the loss function is to promote independence between the latents \citep{Higgins2017}. This encourages the model to find a \textit{disentangled} latent space, where independent factors of variation in the density profiles are captured by different, independent latents. Here, independence is intended in terms of both linearly and non-linearly uncorrelated variables. Although the KL term encourages the network to find disentangled latent parameters, it does not strictly impose disentanglement; we therefore use mutual information (MI) to quantify more strictly the level of disentanglement between the latents.

\subsection{Mutual information}
\label{sec:MI}
In this work, the uses of MI are twofold: (i) to evaluate the degree of disentanglement between the different latents, and (ii) to interpret the latents in terms of their physical information content. MI is a well-established information theoretical measure of the correlation between two random variables. In contrast to linear correlation measures such as the $r$-correlation, MI captures the full (linear and non-linear) dependence between two variables. In other words, the MI between two variables is zero if and only if these are statistically independent.
Mathematically, the MI between two continuous variables $X$ and $Y$ with values over $\mathcal{X} \times \mathcal{Y}$, $I (X, Y)$ is defined as:
\begin{equation}
    \mathrm{MI} (X, Y) \equiv \int_{\mathcal{X} \times \mathcal{Y}}  p_{(X, Y)}(x, y) \left[ \ln{\frac{p_{(X, Y)}(x, y)}{ p_{X}(x) p_{Y}(y)}} \right] {\rm d}x\, {\rm d} y \ ,
    \label{eq:MI}
\end{equation}
where $p_{(X, Y)}$ is the joint probability density distribution of $X$ and $Y$, and $p_X$ and $p_Y$ are their marginal distributions, respectively. MI as defined by Eq.~\eqref{eq:MI} is measured in natural units of information (nats). We make use of the publicly available software GMM-MI \citep{Piras2023}, which performs density estimation using Gaussian mixtures and additionally provides MI uncertainties through bootstrap resampling. We refer the reader to \citet{Piras2023} for further descriptions of MI and their estimator. Our goal is that different latent variables  describe independent factors of variation in the density profiles, implying that the amount of shared information amongst the latents (i.e. their MI) should be negligible. We take MI $\sim \mathcal{O} (10^{-4})$ nats to be a reasonable threshold for disentanglement.

\subsection{Training the IVE}
\label{sec:trainingIVE}

We constructed two sets of training data -- one for TNG100 and the other for TNG100-Dark -- as follows. For each simulation, we split the parent haloes into three sets: $50\%$ for training, $20\%$ for validation and $30\%$ for testing. We then randomly selected 64000/960/6400 subhaloes from the parent haloes and set them aside for training/validation/testing, respectively. We found that increasing the training set size further did not improve the accuracy of the model. The training set was used to optimize the parameters (weights and biases) of the IVE. The validation set does not directly enter the training process of the algorithm, but was used for model selection, i.e., to select the best-performing model amongst the range of possible hyperparameters, and to determine the stopping point for training. 

The training set was sub-divided into batches, each made of $32$ samples. Batches were fed to the network one at a time, and each time the IVE updates its parameters according to the samples in that batch. Training was done using the \texttt{AMSGrad} optimizer \citep{Reddi2019}, a variant of the widely-used \texttt{Adam} optimizer \citep{Kingma2014}, with a learning rate of $10^{-4}$. Early stopping was employed to interrupt the training at the epoch where the validation loss reaches its minimum value.

\begin{figure*}
    \centering
	\includegraphics[width=0.95\textwidth]{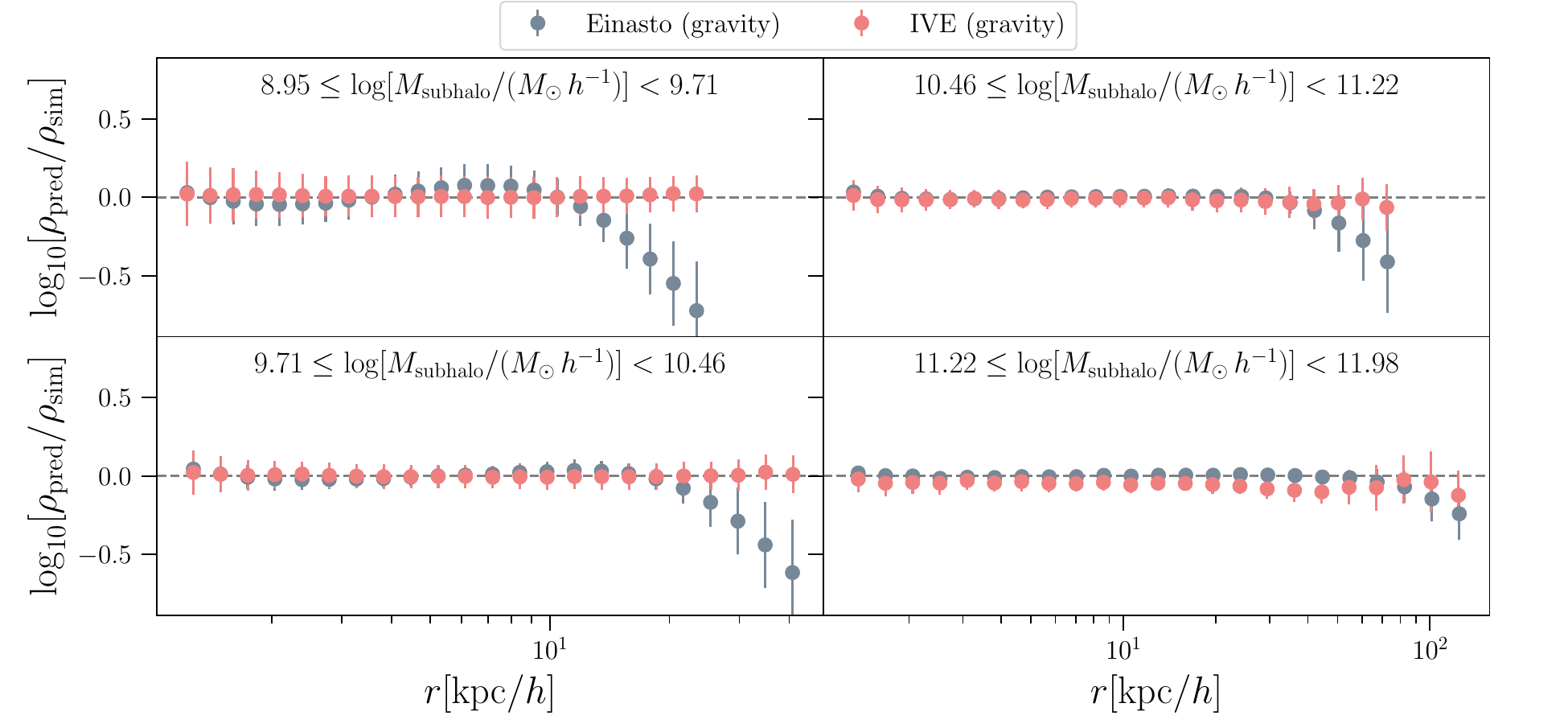}
    \caption{Mean and standard deviation of the residuals $\log_{10}[\rho_\mathrm{pred}/\rho_\mathrm{sim}]$ as a function of radius, where $\rho_\mathrm{sim}$ is the ground-truth profile measured from the simulations and $\rho_\mathrm{pred}$ is the predicted profile for either the Einasto (grey) or the IVE model trained on TNG100-Dark (coral). Each panel shows the residuals for subhaloes of four different mass bins. The radius value on the $x$-axis is given by the mean value of $r$ amongst the subhaloes in each radial bin.}
    \label{fig:pred}
\end{figure*}

We started with the task of finding the underlying dimensionality of the density profile outputs. To do so, we trained four different IVE models with an $L$-dimensional latent space where $L = 2$, $3$, $4$, and $5$, respectively. For this task specifically, we focused primarily on training the model to achieve the best accuracy; thus, we set $\beta$ in Eq.~\eqref{eq:loss} to a very small value, $\beta = 10^{-8}$. In practice, this is equivalent to setting $\beta = 0$. Note that $\beta$ is fixed to $10^{-8}$ only for this initial task of finding the underlying dimensionality of the density profile outputs. We compared the accuracy of the four models and found that the accuracy saturates when $L \geq 3$. We concluded that the underlying dimensionality of subhaloes density profiles is $3$ (see Appendix~\ref{sec:latentdim} for more details). 

Having found the underlying dimensionality, we then proceeded with the next task of achieving both accurate predictions and a disentangled set of latent parameters; here, $\beta$ is not fixed to a single small value but rather varied to achieve the best accuracy-to-disentanglement trade-off. We trained the IVE with $L=3$ (and $L=4$ as a sanity check) latent dimensionality and varying $\beta$ values, aiming to achieve simultaneously highest accuracy and and lowest KL divergence\footnote{We remind the reader that lowest KL divergence corresponds to highest disentanglement amongst latents} in Eq.~\eqref{eq:loss}. In practice, we optimized both the learning rate and $\beta$ via cross-validation. At the end of this process, we were left with two best-performing IVE models, one trained on TNG100 subhaloes and the other on TNG100-Dark ones.

\section{The predicted profiles of subhaloes}
\label{sec:pred}

We compare the predicted profiles of the IVE trained on TNG100-Dark (and on TNG100) to the ground-truth profiles measured from the simulations of individual subhaloes from the testset. The IVE predicts three latent distributions for each subhalo, and a predicted density profile given a randomly sampled latent vector from the latent space. Fig.~\ref{fig:pred} shows the residuals $\log_{10}[\rho_\mathrm{pred}/\rho_\mathrm{sim}]$ where $\rho_\mathrm{pred}(r)$ are the predicted profiles of the IVE (in coral) and $\rho_\mathrm{sim}(r)$ are the ground-truth profiles (in coral). The four panels show the residuals for four different mass bins of subhaloes.
The $x$-axis gives the mean value of $r$ in each radial bin, while the $y$-axis shows the mean and standard deviation of the residuals.

We compare the IVE model to the Einasto density profile \citep{Einasto1965}, a widely-used analytic fitting formula for subhalo density profiles,
\begin{equation}
\rho (r) = \rho_s \exp \left\{ -\frac{2}{\alpha} \left[ \left( \frac{r}{r_s}\right)^\alpha -1 \right] \right\},
\end{equation}
where $r_s$ and $\rho_s$ are the scale radius, defined as the radius at which $\mathrm{d} \ln \rho/\mathrm{d} \ln r =-2$,  and the characteristic density, respectively. The parameter $\alpha$ is a shape parameter that regulates a smooth, gradual transition between the two asymptotic profile slopes of $-1$ and $-3$. We fitted the Einasto formula to each subhalo's profile over the same radial bins used to train the IVE model, by minimizing the expression:
\begin{equation}
\Psi^2 = \frac{1}{N_\mathrm{bin}} \sum_{i=1}^{N_\mathrm{bin}} \left[ \log_{10} \rho_\mathrm{sim, i} - \log_{10} \rho_\mathrm{fit, i}\right]^2,
\end{equation}
where $\log_{10} \rho_\mathrm{sim, i}$ and $\log_{10} \rho_\mathrm{fit, i}$ are the simulation's ground-truth data and the Einasto fitted density profile in radial bin $i$. This expression minimizes the rms deviation between the subhaloes' binned $\rho(r)$ and the Einasto profile, assigning equal weight to each bin. Fig.~\ref{fig:pred} shows the residuals between the Einasto predicted profile and the simulation ground-truth in grey.

\begin{figure*}
    \centering
	\includegraphics[width=0.9\textwidth]{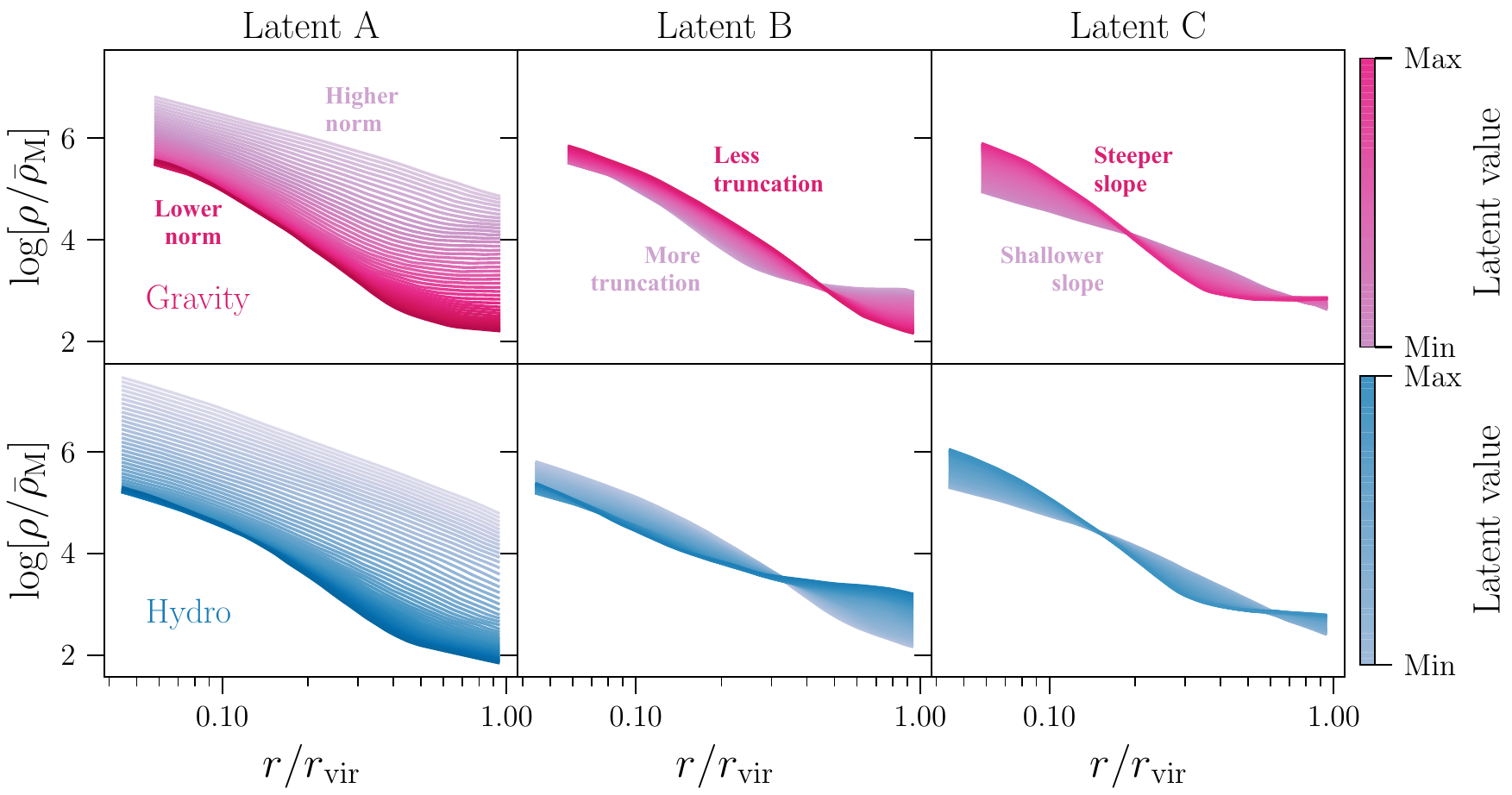}
    \caption{Variations in the predicted density profile of a given subhalo when systematically varying the value of one latent, while keeping the others fixed. Each panel from left to right varies latent A, B or C, respectively; the top and bottom panels show the predictions of the IVE trained on TNG100-Dark and on TNG100, respectively. The first latent describes the normalization of the profile; the second the shape of the profile in the outskirts around the pivot scale $r \sim 0.8\,r_{\rm 200m}$; the third the shape of the inner profile around a smaller pivot scale $r\sim 0.2 \, r_{\rm 200m}$. The inclusion of baryonic effects does not vary the number of degrees of freedom in the model, and their qualitative impact on the profile.}
    \label{fig:varylatent}
\end{figure*}

Figure~\ref{fig:pred} demonstrates that the IVE returns profile predictions that are unbiased and accurate. When compared to Einasto, we find a similar accuracy for the inner radial scales for all mass bins we consider. At intermediate and larger radii that approach the subhaloes' virial radii, and especially for less massive subhaloes, the IVE performs significantly better than the Einasto profile at modelling the total density profile. 
The large decline in accuracy in the Einasto fit at $r \to r_{\rm vir}$ is due to the fact that its functional form is unable to simultaneously fit the subhalo truncation and the outer density plateau; the Einasto profile can only provide a reasonable fit to the density profiles of bound subhalo particles in N-body simulations \citep{Springel2008}. The IVE is instead able to correctly capture the density structure in the outskirts of the subhaloes, thus returning unbiased predictions with similar errors as on smaller radial scales.
As we will show in Sec.~\ref{sec:latents}, tidal truncation induces a first-order effect on the density profile which cannot be ignored. We therefore conclude that, although the Einasto profile can provide reasonable fits to isolated field haloes, it is not a sufficiently good model for subhaloes which experience considerable tidal truncation. The IVE on the other hand provides us with a flexible model that can easily adapt to the corresponding effects using a minimal set of disentangled latent parameters.

We find similar results when training (and testing) the IVE on the total density profiles of subhaloes (including stars, gas and black holes as well as cold dark matter) from the (hydrodynamical) TNG100 simulation; we show the corresponding results in Appendix~\ref{sec:hydrores}. We find that the tidal truncation in the profiles is less pronounced than in the gravity-only simulation due to the presence of baryons, as pointed out in previous work \citep{Dolag2009, Romano2009}. Previous studies also showed that the number of subhaloes in the vicinity of the halo centre substantially decreases in hydrodynamical simulations compared to gravity-only ones; this is because baryonic processes associated with the presence of a large galaxy in the halo centre enhance tidal disruption and destruction by tidal shocking \citep{DOnghia2010, Richings2020}. Since subhaloes closer to the halo centre are also the most tidally disrupted, this effect also points to a population of subhaloes in hydrodynamical simulations that is less tidally disrupted than in gravity-only simulations. As a result, the discrepancy between the Einasto and the IVE model becomes smaller than in the TNG100-Dark case at large radii, albeit it is still present. Additionally, the performance of the Einasto model worsens at intermediate radii, especially for low-mass haloes; as previously mentioned, the flexibility of the IVE allows for significantly better predictions on these scales compared to Einasto. We additionally investigated whether or not more flexible profile fitting functions such as the generalized NFW profile \citep[gNFW -][]{Zhao1996, Freundlich2020} yield better descriptions of the subhalo density profile than Einasto. We find that the gNFW profile yields very similar fits to Einasto in the subhalo outer region and mild improvements in the inner slope.

\subsection{The effect of the latents on the predicted profiles}
In addition to predicting the density profile, the IVE also generates the best-fitting latent distributions associated with each subhalo. To gain further intuition on the profiles modelled by the IVE, we show how the predicted profiles vary as a function of the three latent variables.

Figure~\ref{fig:varylatent} shows the impact of each latent on the predicted profile of an individual subhalo. In each panel, we show the predicted density profiles as we systematically vary the value of one latent, while keeping the others fixed to the mean of their respective Gaussian distributions. The top and bottom panels show the predictions of the IVE trained on TNG100-Dark and TNG100 subhaloes, respectively. The three latents are denoted A, B and C, where the ordering is based on the amount of information each latent captures about the final profile (this will be quantified using MI in Sec.~\ref{sec:latents}).

Latent A captures primarily the normalization of the density profile since varying the latent value shifts the profile in the vertical direction. Latent B determines the shape of the outer profile on the largest radial scales out to the virial radius; in particular, higher/lower latent values produce a steeper/shallower profile in the outer region around a pivot $r\sim 0.6 \, r_{\rm 200m}$. The location of this pivot can change depending on the exact values of latent A and B, but it is always larger than the pivot observed for latent C. The influence of the latent on the profile is similar to the effect of tidal stripping which `truncates' the profile at a radius smaller than the virial radius. Here, this `truncation' effect manifests through a flattening of the profile where the spherically averaged density reaches the background value at radii smaller than the virial boundary. In other words, the smaller the radius at which the profile starts flattening, the larger the mass loss due to the effect of tidal stripping. Latent C primarily affects the steepness (or shape) of the inner part of the profile: higher/lower values of the latent produce a steeper/shallower slope in the inner region. For this particular subhalo, the slope varies around the pivot point $r \sim 0.2\, r_{\rm vir}$, although note that the exact location of the pivot can change depending on which values of A and C are kept fixed. 
The comparison between TNG100-Dark (top panels) and TNG100 (bottom panels) subhaloes shows that the inclusion of baryonic effects does not qualitatively change the meaning of the three latents discovered by the IVE; the latents' impact on the final profile is qualitatively similar to that of the TNG100-Dark latents.

Our results show that, compared to field haloes which can be described by two-parameter models such as the NFW and Einasto (at fixed $\alpha$) profiles, subhaloes require three parameters to model their density profiles up to the virial radius. Two of those are similar to the two needed for field haloes; that is, a normalization and an inner slope parameter similar to the mass and concentration parameters of the Navarro-Frenk-White (NFW) profile \citep{NFW}. Subhalo profiles require an additional third parameter which controls the slope of the outer region where the subhalo experiences mass loss due to a combined effect of tidal stripping, dynamical friction and tidal heating. The same degrees of freedom are required (and sufficient) to describe the total density profiles of subhaloes in both gravity-only and hydrodynamical simulations.

Interestingly, we find that all latents have a non-negligible influence on the profile over the entire radial range. For example, although latent C primarily models the outer profile it also has a non-negligible impact on the amplitude of the profile in the inner region; this amplitude modulation cannot be absorbed into latent A but is instead strictly tied to the physical effect captured by latent C. This in turn implies that one cannot model the effect of tidal stripping based on the outer profile in isolation, as it modulates the entire profile in a non-trivial way. Similarly, as the normalization of the profile is increased via modifications to latent A, the slope of the outer profile also changes. In particular, subhaloes with the highest normalization tend to have a steeper outer slope which differs from the environmental background density; this reflects the fact that high-mass haloes (which have the highest normalization) tend to be less tidally stripped than low-mass (or low normalization) subhaloes.

\section{Interpreting the latents with mutual information}
\label{sec:latents}

We now move to a more quantitative investigation of the information content of the latent parameters, and their relation to the physics of the formation history of subhaloes. To do so, we again make use of MI to quantify the amount of information contained within the latent parameters about other quantities of interest. We start with the MI between each latent and the subhalo density profile. This MI measure provides a complementary approach for interpreting the latent space compared to varying one latent at a time as in Fig.~\ref{fig:varylatent}. The latter shows how the predicted profiles depend on any given latent, conditioned on fixed values of the other latents; the mutual information reveals a more global dependency between latents and ground truths, sensitive to variations in the profiles from all factors simultaneously. In other words, the mutual information tells us how much variation in the ground-truths is captured by each latent at any given radial scale.

We denote the $n$-th latent as $z_n$ where $n=\{$A, B, C, D$\}$, and the ground-truth density in radial bin $i$ as $t_i$; their MI is given by
\begin{equation}
\mathrm{MI}(t_i, z_n) = \int_{t_i} \int_{z_n} p(t_i, z_n) \log \left[ \frac{p(t_i, z_n)}{p(t_i) p(z_n)} \right] {\rm d}t_i\, {\rm d}z_n,
\end{equation}
where $p(t_i, z_n)$ is the joint probability density function between $t_i$ and $z_n$. We make use of the publicly-available software GMM-MI \citep{Piras2023} as mentioned in Sec.~\ref{sec:MI}.

\begin{figure}
    \centering
    \includegraphics[width=\columnwidth]{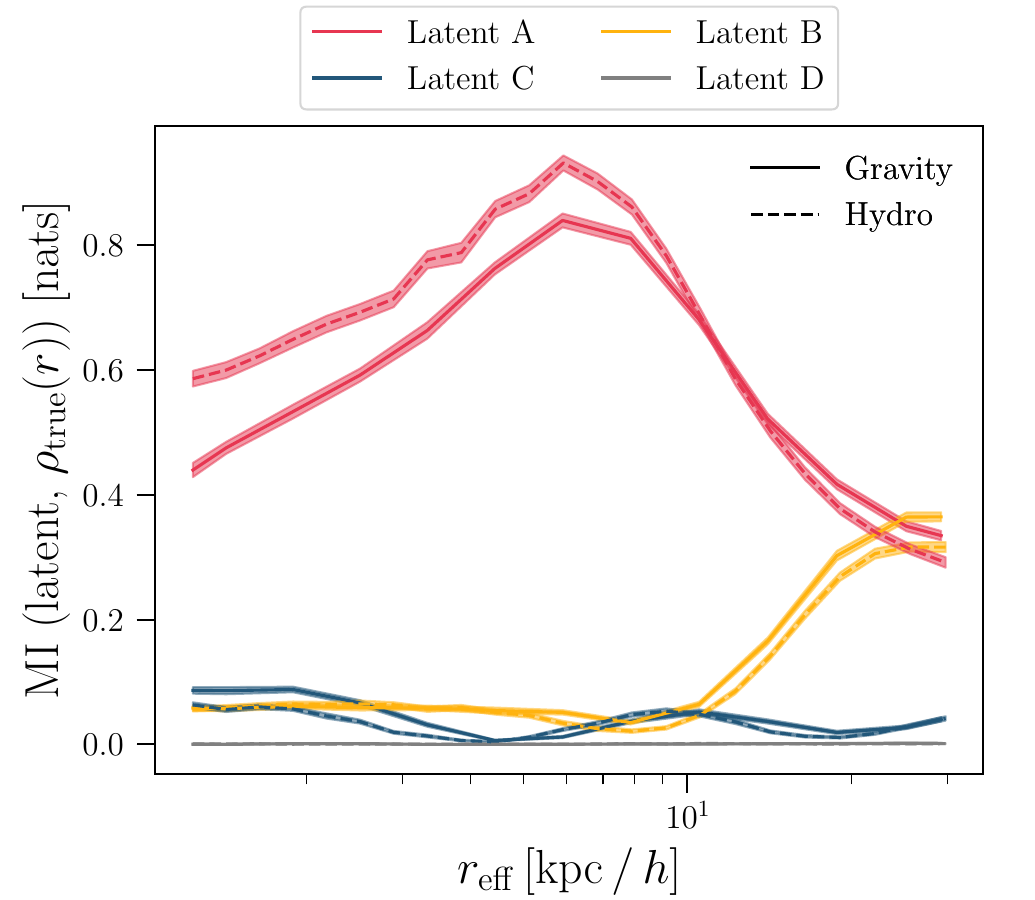}
    \caption{Mutual information (MI) between each latent parameter and the ground-truth spherically averaged density in every radial bin. The variable on the $x$-axis, $r_\mathrm{eff}$, indicates the mean radius in each radial bin across all test set subhaloes. The results for the gravity-TNG100 (solid) and the hydrodynamical one (dashed) are qualitatively consistent.}
    \label{fig:MI_latents}
\end{figure}

Figure~\ref{fig:MI_latents} shows the MI between each latent and the ground-truth density profiles in every radial bin, for both TNG100-Dark (solid) and TNG100 (dashed) subhaloes. While our main results are based on a 3-dimensional latent space model, we present here the results for a 4-dimensional latent space model; this enables us to demonstrate the (non-)impact of an additional fourth latent dimension. Latent A encodes the largest component of variability in the density profiles throughout the entire radial range. Its MI with $\rho_{\rm true}(r)$ increases starting from the smallest radius up to $r_{\rm eff} \sim 5 \, h^{-1} \mathrm{kpc}$ (which on average corresponds to $r \sim r_{\rm vir}/2$), and decreases again all the way to $r_{\rm eff} \sim 12 \,h^{-1} \mathrm{kpc}$ (which on average corresponds to $r \sim r_{\rm vir}$). This means that latent A captures variations in the profile on all radial scales, and it carries most information about the profile shape at intermediate radial scales. 

Fig.~\ref{fig:MI_latents} shows that latent B and C capture less amount of information than latent A, but are nevertheless required for an accurate description of the subhalo profiles. Latent B captures primarily information about the profile in the outskirts at scales approaching the virial radius, where the MI with $\rho_{\rm true}(r)$ reaches values comparable to the MI of latent A. This means that the two latents capture a similar level of (independent) variability in the density profiles; in other words, variations in the profile due to normalization and mass loss are comparable at large radii. Latent C shows two peaks in its MI, one in the core of the subhalo ($r_{\rm eff} \leq 5 \, h^{-1}\mathrm{kpc}$) and one around $r_\mathrm{eff} \sim 8 - 10 \,h^{-1} \mathrm{kpc}$. This is consistent with the picture revealed by Fig.~\ref{fig:varylatent}, where latent C induces changes in the profile above and below the pivot scale $r \sim 6 \, h^{-1}\mathrm{kpc}$. However, latent C's MI never exceeds that of either latent for all values of $r$, meaning that its influence on the density profile, while still important,  is sub-dominant compared to the other two latents. 

For completeness, we also show the impact of including an additional fourth dimension to the latent space; the MI between the fourth latent (latent D) and the density profile is $\mathcal{O}(10^{-4})$ which is two to three orders of magnitude smaller than the MI of the other three latents. This confirms that three degrees of freedom are sufficient to model the subhalo density profile, and any additional latent contains no additional independent information about the profiles.

The comparison between the MI curves for the TNG100-Dark and TNG100 subhaloes reveals a similar story to that of Fig.~\ref{fig:varylatent}. The MI between each latent and the profile for TNG100 subhaloes follows the same trend as a function of radial scale as that for TNG100-Dark subhaloes. In both cases, one latent encodes the largest amount of information about the profile on all radial scales, the other two encode information about the profile outer shape and inner shape respectively. Minor differences in the MI values reflect the differences in the subhalo populations of the TNG100-Dark and TNG100 simulations; the variability in the final profiles differs for different subhalo populations, which in turn affects the amount of variability that can be captured by the latents and therefore the MI values too. We conclude that the latents discovered by the IVE when trained to model subhalo profiles from hydrodynamical simulations have similar information content to those describing subhalo profiles from gravity-only simulations.

The mutual information measure not only reveals the underlying dimensionality but also a `hierarchy' of latents ordered by their information content similar to the principal components of a Principal Component Analysis (PCA) decomposition. Indeed, the latents can be thought of as the principal components of a `non-linear' PCA decomposition. The normalization latent (latent A) is the most important variable which describes the majority of the variance in the profile; the next most important latent is that capturing the outer profile (latent B) affected by tidal stripping; lastly, latent C captures the most subdominant information about the shape of the inner profile. The hierarchical nature of the latent space is achieved naturally through the disentanglement constraint, without the need to design specific layers or impose additional constraints to the loss function \citep{ho2023informationordered}.

\subsection{Relation to the subhaloes' mass accretion history}
\begin{figure}
    \centering
	\includegraphics[width=0.49\textwidth]{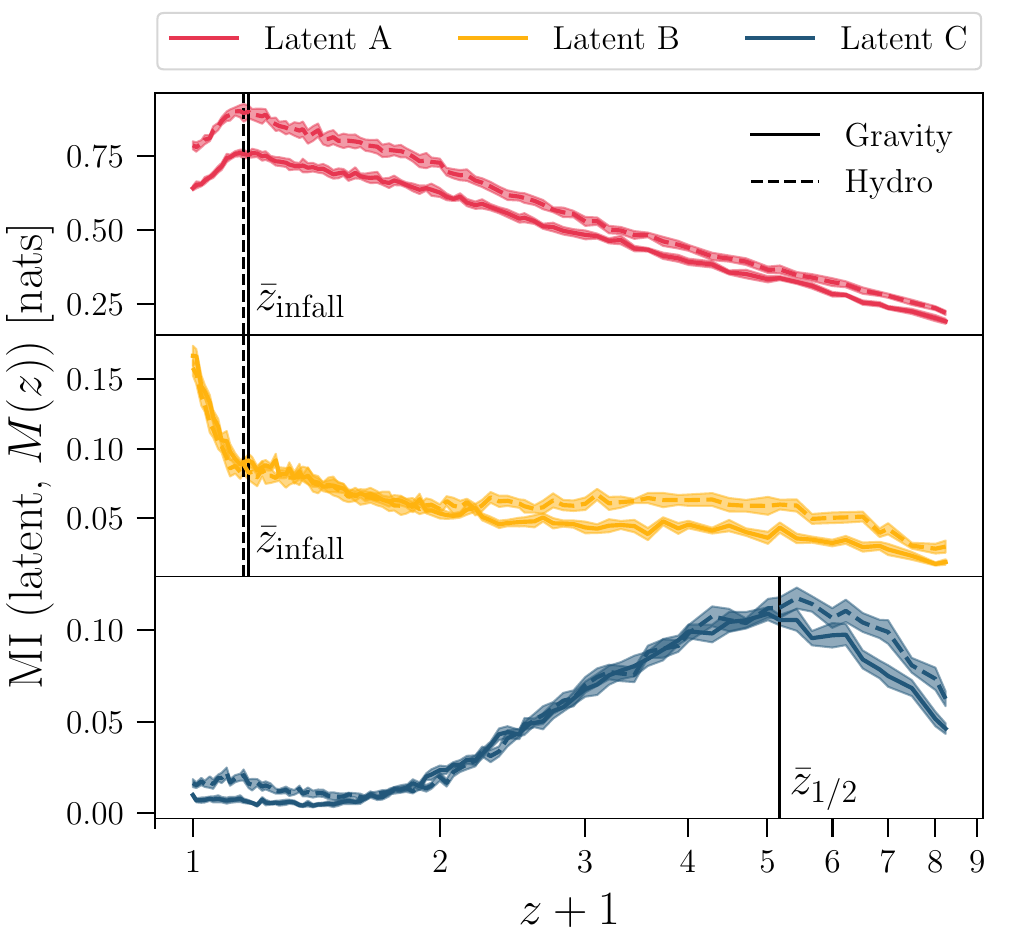}
    \caption{MI between the latents and the subhalo mass accretion histories $M(z)$; top panel shows latent A, middle panel latent B and lower panel latent~C. Solid and dashed lines indicate results for the TNG100-Dark and TNG100, respectively. $z_\mathrm{infall}$ denotes the time at which, on average, the subhaloes fall inside the main parent halo, while $z_{1/2}$ is the redshift at which, on average, the subhaloes accrete half of their present-day mass. Latent A captures the build-up of mass onto the subhalo \textit{before} the average infall time, while latent B probes the subhalo mass loss \textit{after} the average infall time. Latent C captures the early formation history of the subhalo, and peaks at the average subhalo formation time. The latents are connected to the physics of the subhaloes' formation history, despite no information about the latter has been provided during training.}
    \label{fig:miMz}
\end{figure}
The $z=0$ density profiles of substructures (and other cosmic objects) are determined by the complex, non-linear evolutionary history of the subhaloes. The latents, which contain all the information required to model the subhalo profiles, must therefore also be connected to the subhaloes' history. Similar to the work of \citet{LucieSmith2023}, we compute the MI between the latents and the mass accretion histories of the subhaloes. We emphasize that the network does not have access to any information about the subhaloes' mass accretion histories during training — the latents were generated given information about the 3D density field at $z=0$ only.

The mass accretion histories follow the mass of the main progenitor of the subhalo as a function of time. This is constructed from the merger trees which allow one to link the properties of simulated subhaloes across snapshots and identify progenitors and descendants of each subhalo. The \emph{main} progenitor of each subhalo is defined as the one with the ‘most massive history’ behind it \citep[see][]{DeLucia07}, among those that share particles with the target subhalo. The main progenitor is thus not simply the most massive one, removing the arbitrariness, especially in cases when the two largest progenitors have similar masses.

Figure~\ref{fig:miMz} shows the mutual information between the latents and the mass of the subhaloes as a function of time; the three panels show the results for the three different latents. The continuous and dashed lines show the MI results for TNG100-Dark and TNG100, respectively. In the case of latent A (top panel), the MI increases steadily with mass as a function of redshift. This means that latent A is increasingly sensitive to the buildup of mass onto the subhalo as a function of time. The MI peaks at the time where, on average, the subhaloes fall inside the main parent halo -- we call this $\bar{z}_\mathrm{infall}$. This means that latent A is primarily sensitive to the buildup of mass prior to the infall time. After infall, the MI between the latent and $M(z)$ decreases, meaning that the latent does not capture information about those post-infall physical processes that affect the resulting subhalo mass. 

\begin{figure}
    \centering
	\includegraphics[width=\columnwidth]{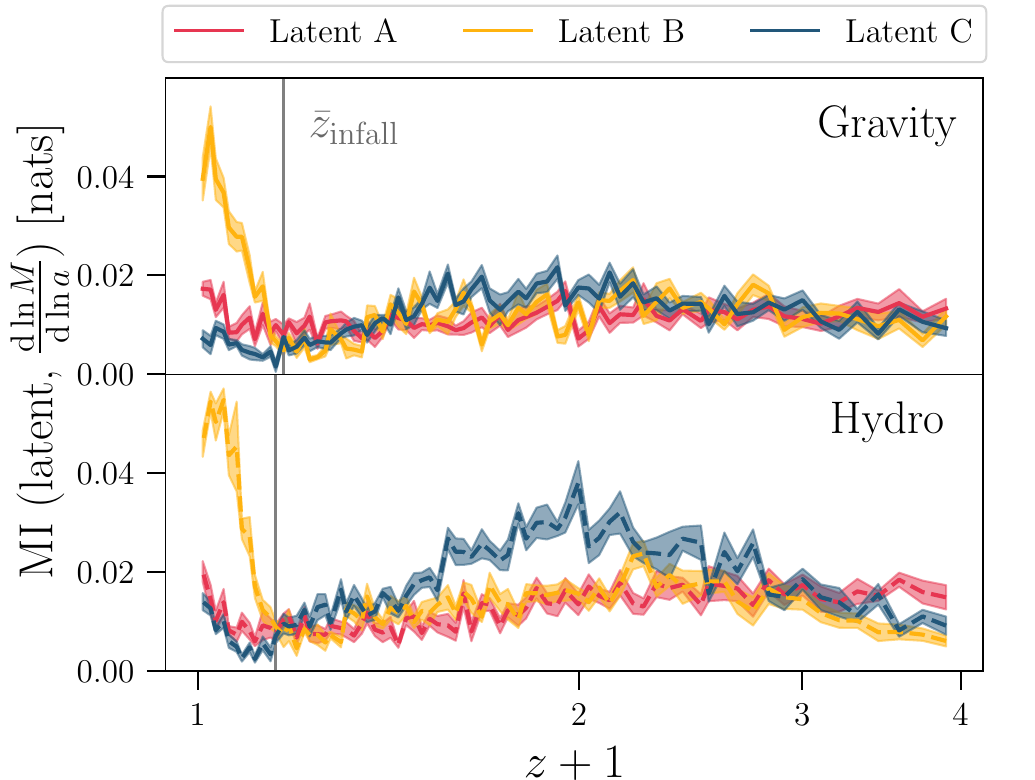}
    \caption{MI between the latents and the subhaloes' mass accretion rate, $\mathrm{d}\ln M(a)/\mathrm{d}\ln a$, for TNG100-Dark (top) and TNG100 (bottom) subhaloes. $z_\mathrm{infall}$ denotes the time at which, on average, the subhaloes fall inside the main parent halo, and it is indicated by a vertical solid (dashed) line for the gravity (hydro) IVE. Latent C, which captures the amount of tidal disruption in the subhalo outskirts, is sensitive to the rate of change of subhalo mass post-infall.}
    \label{fig:midMdz}
\end{figure}

On the contrary, latent B is only mildly sensitive to the formation history prior to infall time; this is shown by the approximately flat (but non-zero) MI between the latent and $M(z)$ for most of the redshift range (middle panel of Fig.~\ref{fig:miMz}). However, the MI starts increasing sharply after $\bar{z}_\mathrm{infall}$, implying that the latent becomes increasingly sensitive to the subhalo mass after infall time. This reveals that latent~B captures information about the impact on the density profiles of physical processes that happen after the subhaloes' infall onto their main host. This is entirely consistent with the picture revealed by Figures~\ref{fig:varylatent} and~\ref{fig:MI_latents} in which latent B captures the shape of the outer profile that is driven by the amount of tidal stripping experienced by the subhalo since it entered its main parent halo. This result is in line with the long-studied effect of tidal stripping in subhaloes from numerical simulations \citep{Diemand2007, Diemand2008, Springel2008, Ghigna2000, Angulo2009, Nagai2005, Nadler2018}, and confirms the ability of the IVE to find physically meaningful latent parameters. To summarise, the IVE found two degrees of freedom that are sensitive to the pre-infall and post-infall accretion history respectively -- all this without any information about the dynamical history of the subhaloes during training. Latent C on the other hand is sensitive to the early-time formation history and peaks at the redshift where, on average, the subhaloes have accreted half of their present-day mass (we denote this $\bar{z}_{1/2}$). This is consistent with latent C being equivalent to the NFW concentration; the latter also affects the shape of the density profiles in the inner region, and it is well-known to carry information about the halo formation times \citep{Wechsler2002, Ludlow2013}.

We find that the physical interpretation of the latents in terms of the subhaloes' formation history is the same for subhaloes from TNG100 and from TNG100-Dark. This means that the independent degrees of freedom in the density profiles with or without including baryonic effects are similarly connected to the formation history of the subhaloes; in other words, baryonic effects do not qualitatively change the physical meaning of the latent parameters. The only differences lie in the exact MI values of each latent, and $M(z)$, which differ because the underlying population of subhaloes naturally differs slightly between TNG100 and TNG100-Dark.

We further test our physical interpretation of the latents by also showing the MI between the latents and the mass accretion rate of the subhaloes in Fig.~\ref{fig:midMdz}. The mass accretion rate is given by $\mathrm{d}\ln M(a)/\mathrm{d}\ln a$ and it is estimated by taking a finite difference between the subhalo mass in subsequent timesteps in the subhaloes' merger trees. The mass accretion rate is a noisier estimate of the formation history of a subhalo compared to the (cumulative) $M(z)$, which explains why the MI values in Fig.~\ref{fig:midMdz} are overall significantly lower than the MI values shown in Fig.~\ref{fig:miMz}. We find that latent A and latent C, controlling the normalization and inner shape of the profile, are largely insensitive to the rate of change in mass; their main source of information lies in the cumulative buildup of mass onto the halo up to infall and formation time, respectively. We note however a slight increase in the MI of latent C at $z \sim 1$; this is more pronounced for the hydro case than the gravity-only case, but present in both. This finding is consistent with the work of \citet{LucieSmith2023} for field haloes showing that the inner shape latent (and the NFW concentration) depends on both the halo formation time during the early assembly phase and the later time mass accretion rate. This dual dependence explains the bimodal shape of the MI between the inner latent and the profile in Fig.~\ref{fig:MI_latents}: the early assembly phase determines the shape of the profile in the innermost region of the halo, while the later time mass accretion rate determines that beyond the pivot on scales $r_{\rm eff} \sim 10 \, h^{-1} \mathrm{kpc}$.

Latent B, controlling the outer profile shape, is instead extremely sensitive to the rate of change of mass after the infall time. This further strengthens our interpretation of the latent capturing the amount of mass lost by the subhalo due to tidal stripping. This in turn changes the boundary of the halo in the outskirts, and therefore the steepness of the profile at those larger scales.

\begin{figure*}
    \centering
 \includegraphics[width=0.95\textwidth]{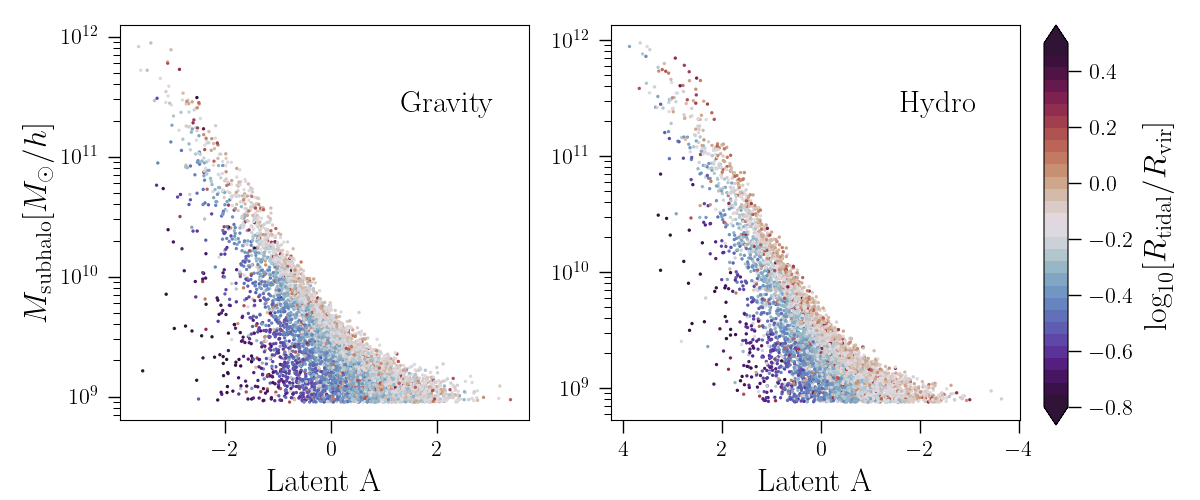}
 \includegraphics[width=0.95\textwidth]{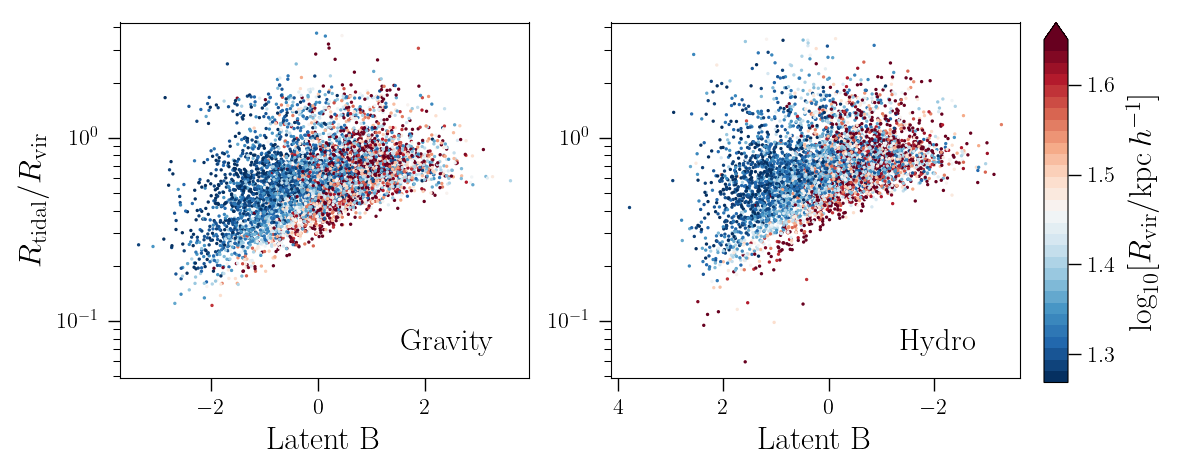}
    \caption{Scatter between physical parameters and latent parameters learnt by the IVE. \textit{Top panels}: Scatter plot between latent A and the mass of the subhalo coloured by $R_\mathrm{tidal}/R_\mathrm{vir}$ for TNG100-Dark (left) and TNG100 (right) subhaloes. Latent A correlates with the mass of the subhalo; the scatter between the two is in turn correlated with the amount of tidal stripping. \textit{Bottom panels}: Scatter plot between latent B and $R_\mathrm{tidal}/R_\mathrm{vir}$, coloured by $R_\mathrm{vir}$ for TNG100-Dark (left) and TNG100 (right) subhaloes. Latent B is correlated with the amount of tidal stripping estimated by $R_\mathrm{tidal}/R_\mathrm{vir}$, albeit with a large scatter largely uncorrelated with $R_\mathrm{vir}$.
}
    \label{fig:scatter_tidalr}
\end{figure*}

\section{Relation between latents and physical parameters}
\label{sec:physicalparams}

We now investigate the relation between the latents and physical parameters typically adopted in the literature related to the study of subhaloes.

We calculate the tidal radius defined as the radius at which the differential tidal force of the host halo is equal to the gravitational force due to the mass of the subhalo \citep{BinneyTremaine87,Tormen1998,Springel2008}. This is commonly assumed to be a good proxy for the subhalo boundary since the expectation is that matter beyond the tidal radius will be removed from the subhalo, thus reducing its mass as it orbits around the host halo. Assuming a subhalo of mass $M_{\rm sub}$ and distance $R$ from the centre of the main halo, the tidal radius can be expressed as
\begin{equation}
R_{\rm tidal}=  R \left(\frac{ M_{\rm sub}}{ \left[ 2- {\rm d}\ln M/{\rm d}\ln r \right] \,M(< R)} \right)^{1/3}  ,
\end{equation}
where $M(< r)$ is the main halo mass within a sphere of radius $r$, and ${\rm d}\ln M/{\rm d}\ln r$ should be evaluated at $R$. The tidal radius is a quantity typically used in the literature to describe the boundary of a subhalo in the presence of tidal stripping. It can therefore be thought of as a proxy for the physics captured by latent B of the IVE.

In Fig.~\ref{fig:scatter_tidalr}, we show scatter plots between various relevant physical parameters and the latent variables. The top panels show the scatter between latent A (controlling normalization) and the mass of the subhalo, coloured by $R_\mathrm{tidal}/R_\mathrm{vir}$, for the TNG100-Dark (left) and TNG100 (right) subhaloes. Note that in the case of the TNG100 subhaloes (right panel), the $x$-axis is flipped to decreasing order to facilitate visual comparisons with the TNG100-Dark case (left panel). The ordering of the latent values does not carry meaning and is arbitrary during training. Latent A correlates with the mass of the subhalo, and the scatter between the two is in turn correlated with $R_{\rm tidal}/R_{\rm vir}$. The latter is a proxy for the amount of tidal stripping experienced by the subhalo, where $R_{\rm tidal}/R_{\rm vir} \sim 1$ is equivalent to no stripping, and $R_{\rm tidal}/R_{\rm vir} < 1$ corresponds to some amount of tidal stripping. This confirms the physical interpretation of latent A: the latent captures the normalization of the profile, which is primarily dependent on the final mass of the subhalo. However, the amount of tidal stripping in the post-infall phase also affects the normalization, which is therefore responsible for introducing scatter in an otherwise direct relation between the mass and the normalization. The latent is able to capture both such dependencies on the profile normalization. The comparison between the TNG100 and TNG100-Dark cases shows the same results for latent A.

The two bottom panels of Fig.~\ref{fig:scatter_tidalr} show the scatter between latent B (controlling the profile outer shape) and $R_\mathrm{tidal}/R_\mathrm{vir}$, coloured by the virial radius $\log R_{\rm vir}$. Similar to the top panels, the $x$-axis in the right panel is flipped to decreasing order to facilitate visual comparisons with the left panel. We find that the relation between the latent and $R_\mathrm{tidal}/R_\mathrm{vir}$ is not so straightforward; while the two are correlated the correlation exhibits a significant amount of scatter. The scatter is not directly related to the mass of the subhalo either, suggesting an intricate non-trivial relation between latent B and commonly used physical parameters (such as the tidal radius and subhalo mass) to describe the tidal stripping. We find a one-to-one mapping between $R_\mathrm{tidal}/R_\mathrm{vir}$ and the distance of the subhaloes from the centre of their main halo (another quantity known to correlate with tidal stripping), which in turn means that the correlation between the latter and latent~B resembles that seen in Fig.~\ref{fig:scatter_tidalr}. We tried various other quantities, such as the mass of the subhalo at infall time, the maximum mass of the subhalo throughout its history, infall redshift, maximum mass redshift; although all these quantites are correlated with the latent to various degrees, we found no clear quantity which could explain the scatter between the latent and $R_\mathrm{tidal}/R_\mathrm{vir}$. We conclude that the IVE latent parameter B, which controls the outer profile shape affected by tidal stripping, does not (and is not expected to) correlate perfectly with any one physical parameter typically adopted in the literature to describe tidal truncation in subhalo profiles. Its information content is a non-trivial combination of the complex physical effects governing the shape of the total density profiles of subhaloes in the outskirts.

\section{Conclusions}
\label{sec:conclusions}

We have presented a novel deep learning model for the density profile of subhaloes with masses $M \gtrsim 7 \times 10^{8} \, h^{-1} \rm{M}_{\sun}$ in the TNG100 simulation. The model consists of an IVE \citep{LucieSmith2022} which predicts the spherically-averaged total density profiles of a subhalo given the raw 3D density field around that subhalo's centre. The IVE first compresses the 3D density field into a low-dimensional, disentangled latent representation, and then maps this latent representation to a prediction for the density profile $\rho(r)$. All the information required by the model to predict the final profile is therefore captured within the latent space. In this way, the IVE not only generates accurate profile predictions but also discovers the underlying degrees of freedom in the subhaloes' profiles through the disentangled latent representation. This serves the dual purpose of understanding and trust: we learn about the underlying physical factors which govern the density distribution within subhaloes while ensuring that the model’s learning is trustworthy and robust. 

We find that a three-dimensional latent representation is required to accurately model the density profiles of subhaloes up to their virial radius.
The first latent captures most of the variability in the density profiles and controls the overall normalization of the profile. This latent is increasingly sensitive to the build-up of mass of the subhalo over time, up to the time when the subhalo falls into its parent halo. The second latent is instead sensitive to the physics of the subhalo formation history after its infall time; in particular, it is driven by the amount of mass loss of the subhalo due to tidal stripping inside the main halo. This latent controls the profile shape in the subhalo outskirts: the profile flattens to the background density at smaller (larger) radii the stronger (weaker) the tidal stripping. To summarise, the IVE disentangles two effects, one controlling normalization and one outer shape, which are sensitive to the pre-infall and post-infall phase of the subhalo formation history, respectively. It should be emphasized that no knowledge about the subhaloes' formation history was provided to the IVE during training.
The third latent resembles the NFW concentration, in that both parameters control the inner shape of the density profile and are sensitive primarily to the mass assembly history of the subhaloes at their half-mass formation time.

We tested the impact of baryonic effects on the density profiles of subhaloes by performing a one-to-one comparison between IVE models trained on TNG100 and TNG100-Dark, respectively. We find that baryonic effects do not have a qualitative impact on our results: the density profiles of subhaloes in both gravity-only and hydrodynamical simulations can be described by the same three degrees of freedom. The physical interpretation of the latents in relation to the subhaloes' formation history is also the same. Our results demonstrate that the modifications to the subhalo profiles induced by baryons can be absorbed with minimal modifications to the parameter space range of the same degrees of freedom as for the pure dark matter case. Our results are limited to the effect of baryonic physics on the density profiles in the galaxy formation prescription adopted by IllustrisTNG; we plan to test our model on different galaxy formation models in future work. 

The evolution of dark matter substructures has been extensively studied in high-resolution cosmological simulations \citep{Kravtsov2004, Springel2008, Dolag2009, Diemand2007}. Our approach for characterizing subhalo density profiles differs from the common approach of manually searching for empirical, analytic fitting functions which often requires introducing many correlated parameters with loose physical motivation \citep{NFW, Einasto1965, Baltz2009, DiCintio2013, Heinze24, Kazantzidis2004}; the range of validity of the different fitting functions and the number of parameters required to account for the relevant physics often remain unclear. Instead, the IVE provides us with a minimal, independent set of ingredients to describe the subhaloes' density profile, which can be directly connected to the physical processes driving the formation history of subhaloes. The IVE latent parameters have a clear physical interpretation in relation to the well-studied subhalo evolution history: the model rediscovers the known correlation between the inner profile and half-mass formation time, and that between the outer truncation and the subhalo mass loss due to tidal stripping. All this was done from $z=0$ inputs alone, without the need to provide any information to the IVE about the subhalo formation histories. The radius at which the outer profile flattens to the background value does not coincide with the well-known tidal radius \citep{Tormen1998, ZavalaFrenk2019}, suggesting that the IVE fitting function is a more complex, non-trivial generalization of the NFW model that goes beyond the truncated NFW model.

One of the main motivations behind this work is the future application to strong lensing analyses. An accurate, robust model for the density profiles of dark matter substructures is key to inferring the correct masses of subhaloes, which are then used to test the $\Lambda$CDM cosmological model. Going forward, we plan to apply the IVE-based density profile model to strong lensing pipelines. The decoder component of the IVE can be used just as it is done with any empirical fitting functions, except that this time we have a neural network replacing an analytic fitting function and the latent parameters as model parameters. Our work shows the promise of using interpretable, deep learning frameworks to model large-scale structure observables in the non-linear regime.

\section*{Acknowledgements}
LLS thanks R{\" u}diger Pakmor for useful discussions. 
LLS and GD thank the organisers of the MIAPbP meeting "Advances in cosmology through numerical simulations", where the ideas for this work were first developed. This research was supported by the Munich Institute for Astro, -Particle and BioPhysics (MIAPbP) which is funded by the Deutsche Forschungsgemeinschaft (DFG, German Research Foundation) under Germany´s Excellence Strategy – EXC-2094 – 390783311. 
GD acknowledges the funding by the European Union - NextGenerationEU, in the framework of the HPC project – “National Centre for HPC, Big Data and Quantum Computing” (PNRR - M4C2 - I1.4 - CN00000013 – CUP J33C22001170001). 

The contributions from the authors are listed below: \textbf{L.L.-S.}: conceptualization; formal analysis; investigation; methodology; software; validation; visualization; writing - original draft, review \& editing. \textbf{G. D.}: conceptualization; investigation; methodology; software; writing -  review \& editing. \textbf{V.S.}: investigation; validation; writing - review \& editing. 

\section*{Data Availability}
The data analysed in this work and the trained IVE models are available upon request to the corresponding author. The entire data of the IllustrisTNG simulations are publicly available at \href{www.tng-project.org/data}{www.tng-project.org/data} and is described in \citet{Nelson2019}.



\bibliographystyle{mnras}
\bibliography{main} 




\appendix

\section{The IVE prediction accuracy}
\subsection{Determining the latent dimensionality}
\label{sec:latentdim}
As mentioned in Sec.~\ref{sec:trainingIVE}, we first focused on finding the underlying dimensionality of the subhalo density profile outputs. We trained four different IVE
models with $L$-dimensional latent spaces, where $L$ = 2, 3, 4, and
5 for the four models respectively. We fixed $\beta=10^{-8}$ in order for the model to focus exclusively on minimizing the predictive accuracy, without the need to disentangle the latent space. We then compared the accuracy of the four models to find the smallest number of latents which saturates the model accuracy.

\begin{figure}
    \centering
	\includegraphics[width=\columnwidth]{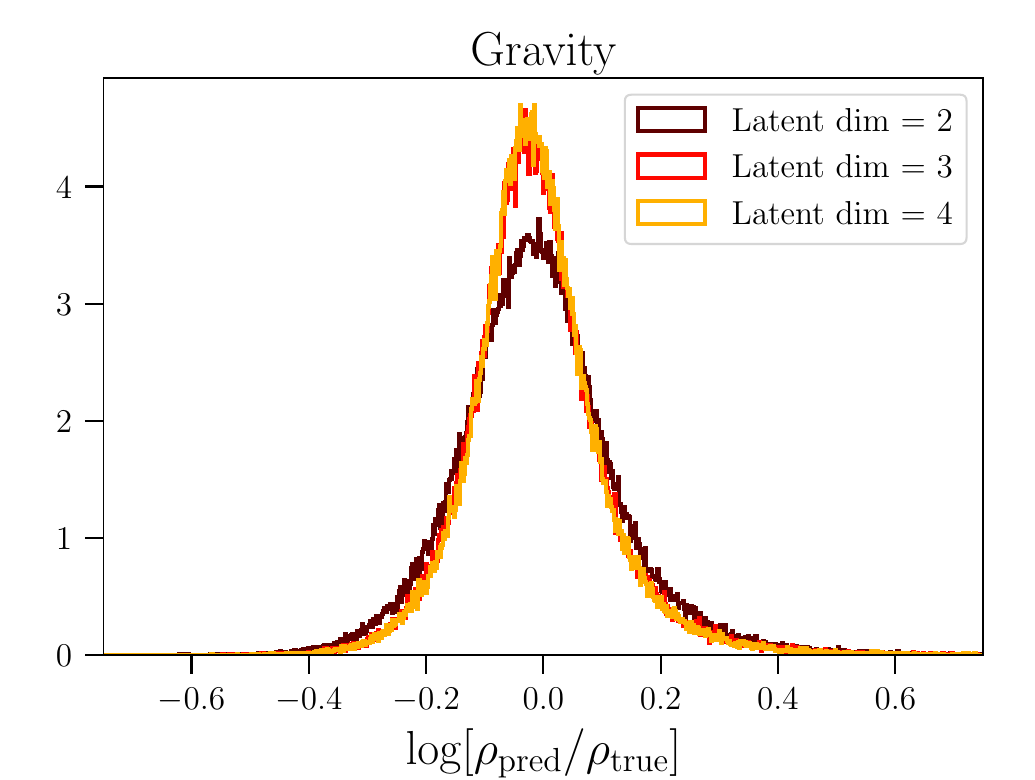}
 \includegraphics[width=\columnwidth]{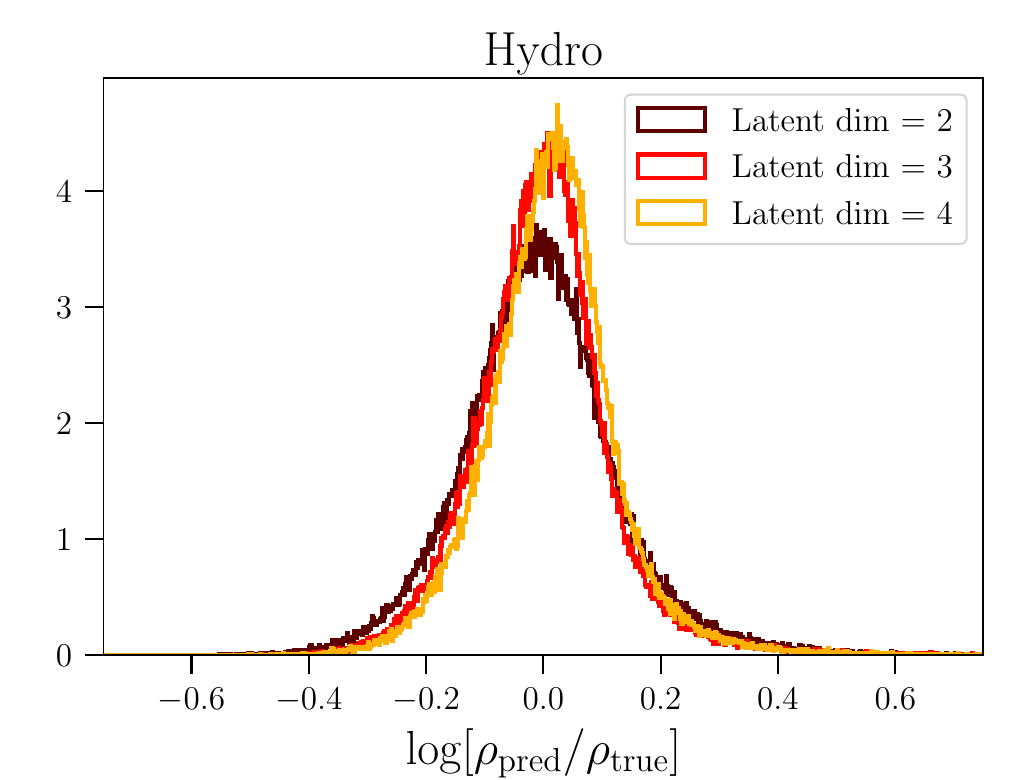}
    \caption{We compare the predictions returned by different IVE models trained using different latent dimensionalities on TNG100-Dark (\textit{upper panel}) and TNG100 (\textit{upper panel}) subhaloes. The histograms show the (log of the) ratio between the predicted and the true density in every radial bin combined.}
    \label{fig:latentdim}
\end{figure}

Figure \ref{fig:latentdim} shows the residuals, $\log [\rho_{\rm pred}/\rho_{\rm true}]
$, for different IVE models trained with different latent dimensionalities, i.e. a 2-, 3- and 4-dimensional latent space. We do not show the results for the 5-dimensional latent dimensionality case to ease visualization. The top panel (bottom panel) shows the case where the IVE is trained on TNG100-Dark (TNG100) subhaloes. We find that the residuals look near-identical when using a 3-dimensional and a 4-dimensional latent space; on the other hand, using a 2-dimensional latent space significantly degrades the accuracy such that the residuals histogram appears broader than histograms of the former two cases. Thus, we conclude that a 3-dimensional latent space is required and sufficient to describe the density profiles of subhaloes.

\subsection{IVE profile predictions in hydrodynamical TNG100}
\label{sec:hydrores}

\begin{figure*}
    \centering
	\includegraphics[width=0.95\textwidth]{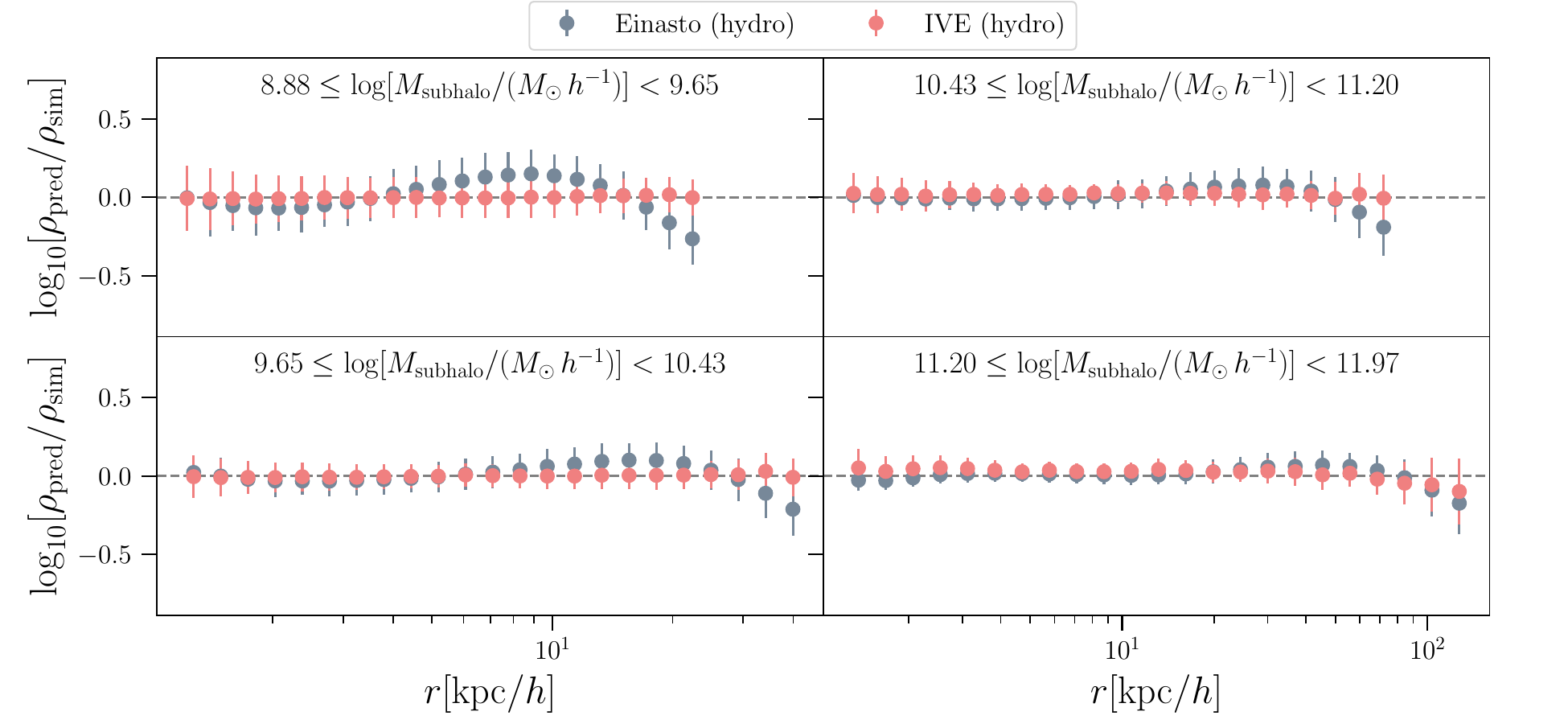}
    \caption{Mean and standard deviation of the residuals $\log_{10}[\rho_\mathrm{pred}/\rho_\mathrm{sim}]$ as a function of radius, where $\rho_\mathrm{sim}$ is the ground-truth profile measured from the simulations and $\rho_\mathrm{pred}$ is the predicted profile for either the Einasto (grey) or the IVE model trained on TNG100 (coral). Each panel shows the residuals for subhaloes of four different mass bins. The radius value on the $x$-axis is given by the mean value of $r$ amongst the subhaloes in each radial bin.}
    \label{fig:predhydro}
\end{figure*}

Figure~\ref{fig:predhydro} shows the residuals $\log_{10} [\rho_\mathrm{pred}/\rho_\mathrm{sim}]$ for the Einasto and the IVE model trained on subhaloes from the hydrodynamical TNG100 simulation, as a function of radius, for four different mass bins in each panel. This is equivalent to Fig.~\ref{fig:pred} but for subhaloes from the hydrodynamical TNG100 (instead of those from TNG100-Dark). The $x$-axis shows the mean value of $r$ in each radial bin out of all subhaloes in that mass bin, while the $y$-axis gives the mean and standard deviation of the residuals for each model. We find that the IVE returns predicitions at similar level of accuracy than in the gravity-only case. Similarly to the TNG100-Dark case, the IVE outperforms the Einasto model; this is especially evident at intermediate radii for the lowest mass range of subhaloes, and at large radii throughout all mass ranges.

As mentioned in Sec.~\ref{sec:pred}, we find that the tidal truncation in the profiles is less pronounced than in the gravity-only simulation due to the presence of baryons. This, in turn, makes the discrepancy between the Einasto model and the simulation profiles less pronounced at large radii, albeit it is still present. On the other hand, we find that the performance of the Einasto model worsens at intermediate radii, especially for low-mass haloes, compared to the pure dark matter case. As previously mentioned, the flexibility of the IVE allows for significantly better predictions on these scales compared to Einasto.


\bsp	
\label{lastpage}
\end{document}